\documentclass[prl,twocolumn,showpacs]{revtex4}

\usepackage{graphicx}
\usepackage{amsmath}

\newcommand{\eq}{\begin{equation}}
\newcommand{\en}{\end{equation}}
\newcommand{\sN}{{\cal N}}
\newcommand{\sT}{{\cal T}}
\newcommand{\sD}{{\cal D}}

\begin{document}

\newcommand{\EXXON}{ExxonMobil Research and Engineering, 1545 Route 22 East, Annandale, NJ 08801, USA}

\title{Coherent Structures in Dense Granular Flows}

\author{Thomas C. Halsey}
\author{Deniz Erta{\c s}}
\affiliation{\EXXON}
\date{\today}

\begin{abstract}

We present a theoretical derivation of a rheology for dense granular flow, based on the process of inelastic collapse of neighboring particles.  This collapse creates regions of correlated motion, which control the viscous behavior of the flow on a large scale.  The result is a rheology that obeys the scaling form observed experimentally by Pouliquen and by the ``G.D.R. Midi" group author. We identify the nature of the constraints imposed by inelastic collapse on the grain-scale motions in the flow; finally, we show using an energy cascade argument that the inelastic collapse need not proceed to the final endpoint in order for the correlations we have identified to build up.

\end{abstract}


\maketitle

Dense granular flows are ubiquitous in nature and in technology \cite{gollub}.  Notwithstanding, robust, microscopically well-founded descriptions of the behavior of these flows across flow geometries have proven elusive \cite{flows}. Several recent developments have hinted at the outline of such a description. In 1999, Pouliquen published a seminal paper performing a scaling analysis of ``chute flows" on an inclined plane, showing that the rheology of these flows was controlled by a single scaling length \cite{po1999}.  These results were reproduced in numerical studies \cite{sier2001}, and a potential explanation of this scaling length as arising from correlations in granular motion was advanced by the authors of this Letter \cite{erha2002,louge}. This scaling length has also been seen experimentally \cite{length}. Most recently, a group author ``G.D.R. Midi" has published a comprehensive comparison of numerical and experimental results on dense granular flows in a number of geometries, including both confined systems and systems with free surfaces \cite{gdrm}.  They argue that for a system made of particles of diameter $d$, of density $\rho$, flowing at a shear rate $\dot \gamma$, at a pressure $P$, the rheology is controlled by the local scaling variable
\eq
I = \frac{\dot \gamma d}{\sqrt{P/\rho}} ,
\label{eq:gdrm}
\en
which figures as a generalization of the Pouliquen result.

In this Letter we extend our earlier work on chute flows to account for Eq.~(\ref{eq:gdrm}) as a scaling variable for dense granular flows.  We present a microscopic picture of the origin of this quantity in the dynamics of the inelastic collapse of the particles in a flow, and we show how to apply these ideas to generate a systematic description of two standard flows: pure shear at constant pressure, and chute flow with a free surface.

It is well known that inelastic collapse is a key phenomenon in many types of granular systems \cite{gold}. Consider two particles brought together by a restoring force $F$ at a normal velocity $v_0$. If the particles have a restitution coefficient $\epsilon$, and they are perfectly rigid so that the collisions are instantaneous, we expect that they will collapse onto one another in a time
\eq
\tau_{nnc} = \frac{2 m v_0}{F} \frac{\epsilon}{1-\epsilon} .
\label{eq:fcollapse}
\en

Real granular particles are not perfectly rigid, and may also have velocity-dependent coefficients of restitution \cite{velocity}.  In practice, inelastic collapse will be controlled by a balance between a time scale $\tau_c$ that measures the duration of collisions and a time scale $\tau_b$ that measures the intervals of ballistic motion between collisions. For perfectly rigid particles $\tau_c \to 0$, but of course $\tau_b \to 0$ as well in the late stages of inelastic collapse. Thus a thorough analysis of inelastic collapse in a flow requires careful specification of the elastic and dissipative mechanisms of particle collision, as well as a clear sense of the flow conditions to which the particle is subject. 

We are first going to finesse this complexity by simply assuming that the time scale $\tau_{nnc}$ to reach the regime in which $\tau_c \gg \tau_b$ is short compared to the inverse shear time $\dot \gamma^{-1}$. Unfortunately, if one takes the limit of infinite rigidity of particles while keeping the coefficient of restitution and shear rate constant, one will never reach this regime.  For this reason, at the end of this Letter we will consider the case in which $\tau_b \gg \tau_c$ always, and show that the results for the flow rheology are similar.

Moving from the case of two particles to the case of a dense granular system at a pressure $P$, it is natural to generalize Eq.~(\ref{eq:fcollapse}) to
\eq
\tau_{nnc} = \frac{\rho d^2 \dot \gamma}{P} f(\epsilon) ,
\label{eq:pcollapse}
\en
where the local pressure $P$ provides the restoring force, and where the unknown function $f(\epsilon)$ accounts for the effect of multiple collisions with other particles in the course of an inelastic collapse event. The varying frequency of collisions needed to maintain the pressure on the grain may cause additional dependence of $f(\epsilon)$ on the local packing fraction, which we have suppressed in this simplified analysis.

For a granular system in motion with a local shear rate of $\dot \gamma$, the time scale $\dot \gamma^{-1}$ sets the time over which the neighbors of a particle are changed by the overall shear flow.  Provided $\dot \gamma \tau_{nnc} \ll 1$, we expect that clusters will form, the motion of whose particles are correlated with one another by this process of inelastic collapse.  The formation of these clusters will be driven by the diffusive process of neighboring particles collapsing onto one another; each such collapse event takes a time of $\tau_{nnc}$ and introduces correlations into the flow on a scale of the interparticle distance, or the particle diameter $d$ in the collapsed state. Thus the length scale of such correlations $l_E$ will be determined by
\eq
l_E = \sqrt{ \frac{d^2}{\tau_{nnc}} \frac{1}{\dot \gamma}},
\label{eq:lE}
\en
where $d^2/\tau_{nnc}$ figures as a diffusion constant for a process occuring over a time scale of $\dot \gamma^{-1}$. We thus obtain
\eq
l_E =  \sqrt{\frac{1}{ f (\epsilon)}\frac{P}{\rho}} \frac{1}{\dot \gamma} .
\label{eq:defe}
\en
Note that there is no explicit dependence here on the particle diameter $d$.

The precise nature of the correlations inside a cluster is somewhat subtle. When particles collapse onto one another, the relative normal velocity of the particles is driven to zero. In addition, the influence of friction will couple the rotational motion of the particles to their transverse velocity; immediately, in the case of infinite friction, or after a finite time, in the case of finite friction. Consider a set of $N$ particles $i$ in three dimensions with velocites $\vec v_i$ and rotational velocities $ \vec \omega_i$ (the latter the axial vector corresponding to the rotation about an axis $\hat \omega$).  We further suppose the existence of pairs of particles $\langle ij \rangle$ that have inelastically collapsed onto one another, and for which there is no relative motion of the surface points in contact (corresponding to frictional locking of the particles.)  Suppose that the vector connecting the collapse pair $\langle ij \rangle$ is $\Delta \vec w_{\langle ij \rangle}$. Then the requirement that the relative motion of the surface points in contact be zero is
\eq
\vec v_i - \vec v_j \equiv \Delta \vec v_{\langle ij \rangle} = \frac{1}{2} (\omega_i +\omega_j) \times \Delta \vec w_{\langle ij \rangle} .
\label{eq:collapse1}
\en
Taking the derivative with respect to time of this constraint yields a constraint for the accelerations of the particles $\vec a_i$ and angular accelerations $\vec \Gamma_i \equiv \frac{d}{dt} \vec \omega_i$
\eq
\Delta a_ {\langle ij \rangle} =  \frac{1}{2} (\omega_i +\omega_j) \times \Delta \vec v_{\langle ij \rangle} +  \frac{1}{2} (\Gamma_i +\Gamma_j) \times \Delta \vec w_{\langle ij \rangle},
\label{eq:collapse2}
\en
where, fortunately, complications owing to the non-Abelian nature of the rotation group do not appear to this order in $d/dt$.  We call states constrained by Eqs.~(\ref{eq:collapse1},\ref{eq:collapse2}) ``gear" states, for the obvious reason that the particles are rolling over one another like gears.

These equations significantly constrain both the motion of the particles and the forces between the particles. If the average coordination number of the collapsed particles is $z$, then Eq.~(\ref{eq:collapse1}) gives three constraints per contact or $3z/2$ per particle. Since each particle has 6 degrees of freedom without contacts, this means that the effective number of degrees of freedom per particle $N_F/N$ is
\eq
\frac{N_F}{N} = \frac{3}{2}(4-z) .
\label{eq:coord}
\en
We thus see that the average coordination $z<4$ in order for the collapsed state to be mobile at all, and Eq.~(\ref{eq:coord}) gives the effective number of degrees of freedom for the collapsed state \cite{edwards}.

The forces are even more completely determined.  The total number of contact forces is exactly the same as the number of constraints on the accelerations given by Eq.~(\ref{eq:collapse2}), with the result that all of these forces are uniquely determined by the contact network and the velocities and angular velocities of the particles.

Of course, the constraint equations on the velocities and accelerations do not hold indefinitely. At a certain point in the flow of the collapsed region, one of two breakdowns in these equations must occur (see Figure 1). The first of these is the collision of two particles, which will initiate a new episode of inelastic collapse, at the end of which the contact network $\{ {\langle ij \rangle} \}$ and the velocities will be altered by the various impulses created in the network of collapsed particles by this collision. The second type of failure is associated with the forces between particles. These are not arbitrary-- the normal forces $\sN$ between granular particles may not be negative, and the tangential forces $\sT$ for particles with finite coefficients of friction $k$ must obey a Coulomb constraint $\sT \le k \sN$. As a packing moves, we expect all forces determined by Eq.~(\ref{eq:collapse2}) to develop smoothly for some finite length of time, until one of these constraints is violated. At this time, a contact will fail and the contact network will change discontinuously.

\begin{figure}[t]
\begin{center}
\includegraphics[scale=0.5]{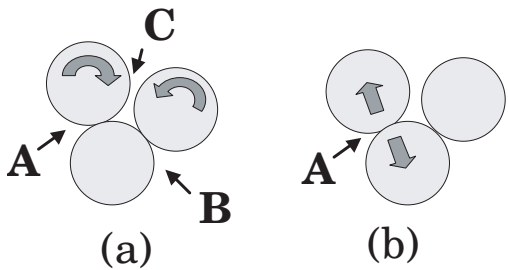}
\caption{
Two means of disrupting the evolution of a collapsed coherent state: a) formation of a new contact at C due to inter-particle collisions--the particles are rolling on contacts A and B, and b) failure of an old contact at A under tension, or due to exceedance of a Coulomb criterion on the tangential force $T \le k N$ for normal force $N$ at a contact.
}
\end{center}
\label{figureone}
\end{figure}

Having elucidated the nature of the correlations in the collapsed region, we return to the role of these regions in controlling the overall rheology of a granular flow. Our fundamental assumption is that the collapse scale $l_E$ plays a role analogous to a Prandtl mixing length in hydrodynamics. Consider first the consitutive equation relating the shear component of the stress tensor $\sigma$ to the local shear rate $\dot \gamma$
\eq
\sigma = \mu \dot \gamma ,
\label{eq:fund}
\en
defining a viscosity $\mu$. We can write this viscosity in terms of a length scale $l_{\mu}$,
\eq
\mu = \rho l_{\mu}^2 \dot \gamma ,
\label{eq:defmu}
\en
effectively defining $l_{\mu}$. So far these manipulations are constrained only by dimensional analysis. We now introduce the physical assumption that we can write
\eq
l_{\mu} = \tilde a l_{E}(1 + \tilde b \frac{d}{l_E} + \cdots) ,
\label{finitesize}
\en
where this relation is written in the form of an asymptotic expansion in $d/l_E$, and $\tilde a$, $\tilde b$ are constants.  This relation is the heart of our approach.  Note that the sign of the first finite size correction $\tilde b d / l_E$ is constrained by the fact that conventional Bagnold scaling would imply that when $l_E \to 0$, $l_{\mu} \to d$ \cite{flows}.  (Although once we reach the limit $l_{\mu} \sim d$ more conventional kinetic theory approaches may be superior to our approach.)

We can now solve for the full shear dependence of $\mu$. From Eqs.~(\ref{eq:defe},\ref{eq:defmu}), we see that
\eq
\frac{\mu \dot \gamma}{P} = f^{-1} (\epsilon) \left (\frac{l_{\mu}}{l_{E}}\right )^2 ,
\label{eq:ratio}
\en
or
\eq
\mu = \frac{\tilde a^2 P}{ f(\epsilon) \dot \gamma}\left ( 1 + 2 \tilde b \sqrt{  f(\epsilon)} \frac{\dot \gamma d}{\sqrt{P/\rho}} + \cdots \right ) ,
\label{viscosity}
\en
which together with Eq.~(\ref{eq:fund}) defines the rheology (excepting more subtle effects such as normal stress differences) This is a Bingham-type rheology, with a yield stress added to a dynamic stress linear in the shear rate, although the pressure-dependences are novel compared to typical Bingham rheologies \cite{rheology}. The alert reader will note the scaling dependence on the parameter $I$ defined by the G.D.R. Midi group (Eq.~(\ref{eq:gdrm})).

Now let us consider the application of this result to some standard flow geometries. Consider first simple shear. In this case we fix the normal force per unit area $F_N$ exerted into a sample by a boundary, as well as the tangential force per unit area $F_T$ exerted at that boundary. Let us presume that we can write the pressure $P$ as $P=(1+K)F_N/3$, with $K \ne 2$ in the case where normal stress differences develop \cite{normal}. This yields
\eq
\frac{F_T}{F_N} = \frac{\tilde a^2 (1 + K)}{ 3  f(\epsilon)}\left ( 1 + 2 \tilde b \sqrt{ f(\epsilon)} \frac{\dot \gamma d}{\sqrt{(1+K)F_N /3\rho}} + \cdots \right ) ,
\en 
showing a linear increase of the shear rate with tangential stress beyond a yield stress proportional to the normal force.  Note that for a cell of size $L$, we must impose $d < l_E < L$ in order for the inelastic collapse approach to make sense; indeed, we expect the flow not to be mobile if $l_E > L$, because collapse onto the wall will become possible on time scales shorter than $\dot \gamma^{-1}$.

Turning to the case of flow down an incline at an angle $\theta$, for which $\sigma = \rho g \sin \theta$, and $P=\frac{1}{3}(1+K) \rho g \cos \theta$, we obtain
\eq
\tan \theta = \frac{\tilde a^2 (1 + K)}{3 f(\epsilon)}\left ( 1 + 2 \tilde b \sqrt{  f(\epsilon)} \frac{ \dot \gamma d}{\sqrt{(1+K)g \cos \theta/3}} + \cdots \right ).
\en
It was shown already in Ref.~\cite{erha2002} that this type of formula accounts well for the phenomenology of chute flows, as expounded in Ref.~\cite{po1999}.

Our assumption that particles are correlated through collapse events of finite duration is somewhat naive--in practice the generation of ``thermal" energy in the state through the impulses generated by the creation of new contacts (see Figure 1) will tend to disrupt contacts. It is therefore instructive to consider in more detail the energy balance in a collapsing packing.  In this way we can actually generalize our results to cases where inelastic collapse need not proceed to the end state of contacts of arbitrarily long duration.

Our approach is to write formulae for the dissipation of energy on a succession of scales, and require that they be equal. Let us first consider the work done on a macro scale.  From continuum theory, the rate at which work is done by the flow is
\eq
\sD_M = \sigma \dot \gamma.
\label{dissm}
\en
Returning to Figure 1, we see that in the collapsed state (which is our highest-level description of this process), the only dissipative process is that indicated in Figure 1a, in which a particle collides with another particle.  Since the particles are part of extended objects of scale $l_{\mu}$, the effective momentum difference in this collision is actually $\Delta p \propto m \dot \gamma l_{\mu}$, so that the rate of dissipation on a time scale where collapse events appear instantaneous is
\eq
\sD_C =  \rho \dot \gamma^3 l_{\mu}^2.
\label{dissc}
\en
Analogously to our discussion above, we can regard this as a definition of $l_{\mu}$, but with the interpretation that this result corresponds to the dissipation associated with a particle collapsing onto an extended object of size $\sim l_{\mu}$.

Now we resolve the flow on finer time and length scales, on which the particles are not fully collapsed onto one another. We suppose that the particles are actually separated by small distances $\delta \ll d$, with ballistic flight times across these distances (between collisions) of $\tau_b$ and collision durations of $\tau_c$.  If $\tau_c \gg \tau_b$, then in practice the packing is collapsed. This is the case that we examined above, which led to the fundamental rheology given by Eq.~(\ref{viscosity}). If $\tau_b \gg \tau_c$, then the packing is somewhat similar to a granular gas; the difference between our approach and normal kinetic theory approaches to motion is that in considering motion on scales $\delta$ relative to the collapsed state, we effectively consider motion relative to the ``gear'' state defined by Eqs.~(\ref{eq:collapse1}-\ref{eq:collapse2}). Normal kinetic theory approaches to flow consider thermal motion as being defined relative to a much simpler laminar flow in determining the ``thermal'' component of velocity \cite{kinetic}.

If there is a constant coefficient of restitution of $\epsilon$, the loss of energy through inelastic ``thermal'' collisions is
\eq
\sD_I = \rho (1- \epsilon )\frac{\delta^2}{\tau_b^3},
\label{dissi}
\en
where the typical mean-free path $\delta \ll d$ in a dense granular system. Clearly we require
\eq
\sD_M= \sD_C = \sD_I.
\label{disseq}
\en

For a non-zero coefficient of friction, the transverse translational and rotational motion of the particles will lock as determined by Eq.~(\ref{eq:collapse1}) after a small number of collisions; we will get an approximately correct result if we assume that the particles lock after one collision, which will indeed be the case for an infinite coefficient of friction. The size of a correlated region $l_E$ is then determined by reprising the diffusive argument leading to Eq.~(\ref{eq:lE}),
\eq
l_E^2 = \frac{d^2}{\tau_b} \frac{1}{ \dot \gamma}.
\label{eq:thermdiff}
\en
Again, we expect Eq.~(\ref{finitesize}) connecting $l_{\mu}$ to $l_E$ to hold. We can write the pressure from ordinary kinetic theory arguments as
\eq
P \propto \frac{1}{d^2} \frac{m \delta}{\tau_b^2},
\en
or
\eq
\frac{P}{\rho} = \tilde B \frac{d \delta}{\tau_b^2},
\label{eq:pressure}
\en
with $\tilde B$ a constant, which completes our system of equations \cite{pressure}.  Note that while the pressure is constrained by the order of magnitude expression Eq.~(\ref{eq:pressure}), the average force between particles on time scales long compared to $\tau_b$ but short compared to $\dot \gamma^{-1}$ is still fixed by the gear equations Eqs.~(\ref{eq:collapse1},\ref{eq:collapse2}).

After some manipulation, it is possible to reduce Eqs.~(\ref{dissm}-\ref{eq:thermdiff},\ref{eq:pressure}) to
\eq
\frac{\sigma}{P} \equiv \frac{\mu \dot \gamma}{P} = \frac{\sqrt{(1-\epsilon)} l_{\mu}}{\tilde B l_E},
\en
which is similar to Eq.~(\ref{eq:ratio}), with the exception that the ratio $l_{\mu}/l_E$ appears to the second power in the former case, and to the first power here. Applying Eq.~(\ref{finitesize}) connecting $l_{\mu}$ to $l_E$, we see that this difference actually amounts at first order in $d / l_E$ only to a factor of two in the shear-rate dependent term in the effective viscosity, as well as changes to the overall constants multiplying the formula. Thus, while the precise nature of the inelastic collapse, and in particular the relative magnitudes of $\tau_b$ and $\tau_c$, do affect details of the rheology, they do not affect the qualitative result.

We are grateful to P.M. Chaikin, N. Mitarai, H. Nakanishi, and O. Pouliquen for illuminating discussions.  T.C.H. is grateful to the Kavli Institute for Theoretical Physics (KITP) for their support;  this research was supported in part at the KITP by the National Science Foundation under Grant No. PHY99-07949.


\begin{thebibliography}{99}
\bibitem{pressure} The pressure formula allows one to rephrase the difference between the two cases analyzed--the collapse case analyzed in Eqs.~(\ref{eq:pcollapse}-\ref{eq:defe}) assumes that the initial velocity before inelastic collapse commences is $v_0 = \dot \gamma d$, while the approach of Eqs.~(\ref{dissm}-\ref{eq:pressure}) can be derived also from Eq.~(\ref{eq:fcollapse}) if the initial velocity is taken to be $v_0 = \delta / \tau_b$.
\end{thebibliography}
\end{document}